\documentclass{ws-ijgmmp}
\usepackage[utf8]{inputenc}

\begin{document}

\markboth{Piyali Bhar}
{Strange star with Krori-Barua potential in presence of anisotropy}

%
\catchline{}{}{}{}{}
%

\title{Strange star with Krori-Barua potential in presence of anisotropy}

\author{Piyali Bhar}
\address{Department of
Mathematics,Government General Degree College, Singur, Hooghly, West Bengal 712 409,
India.\\
\email{piyalibhar90@gmail.com }}

\maketitle

\begin{history}
\received{(Day Month Year)}
\revised{(Day Month Year)}
\end{history}

\begin{abstract}
In present paper a well-behaved new model of anisotropic compact star in (3+1)-dimensional spacetime has been investigated in the background of Einstein's general theory of relativity. The model has been developed by choosing $g_{rr}$ component as Krori-Barua (KB)
{\em ansatz} [Krori and Barua in J. Phys. A, Math. Gen. 8:508,
1975]. The field equations have been solved by a proper choice of the anisotropy factor which is physically reasonable and well behaved inside the stellar interior. Interior spacetime has been matched smoothly to the exterior Schwarzschild vacuum solution and it has also been depicted graphically. Model is free from all types of singularities and is in static equilibrium under different forces acting on the system. The stability of the model has been tested with the help of various conditions available in literature. The solution is compatible with observed masses and radii of a few compact stars like Vela X-1, 4U $1608-52$, PSR J$1614 - 2230$, LMC X $-4$, EXO $1785-248$.
\end{abstract}



\section{Introduction}
The study of relativistic stellar structure is not a new topic, rather it was initiated
more than $100$ years ago by Karl
Schwarzschild {\em et al.} \cite{karl} with the discovery of the universal vacuum exterior solution. In the same year, they also published a paper which described the first interior solution \cite{karl1}. Jeans \cite{jeans} first obtained the model of compact star by choosing pressure anisotropy in self gravitating
objects in the context
of Newtonian gravity. Ruderman \cite{rud} and Canuto \cite{can} first proposed the concept of anisotropy which influenced many researchers to study the fluid sphere in presence of pressure anisotropy by using general relativity. They claim that at the center of the compact star model where the density is beyond the nuclear density, i.e., at the order of $10^{15}~gm.cm^{-3}$, the pressure can be decomposed into two parts one is radial pressure $p_r$ and another is transverse pressure $p_t$ in the perpendicular direction to $p_r$. Bowers and Liang \cite{liang} first generalized the equation of hydrostatic equilibrium for the case local anisotropy. As proposed by Kippenhahn and Weigert \cite{kip} anisotropy may occurs by existence of a solid stellar core or by the presence of a type-3A superfluid, pion condensation \cite{pion}, different kinds of phase transitions \cite{phase}, mixture of two gases (e.g., ionized hydrogen and electrons) \cite{lete}. Spherical
galaxies in presence of anisotropy in the context of Newtonian gravitational theory was obtained by Binney and Tremaine \cite{bin}. Strong magnetic fields also creates anisotropic pressures
inside a compact sphere proposed by Weber \cite{weber}. Dev and Gleiser \cite{dev1,dev2} have shown the effect of pressure
variation on physical properties of a compact star model. Rahaman {\em et al.} \cite{rah} and Bhar \& Rahaman \cite{pbf} also used anisotropic pressure to find the existence of the wormhole in higher dimension in the noncommutativity-inspired spacetime. In both the papers the authors have shown that in presence of pressure anisotropy, the wormhole solutions exist only in
four and five dimensions; however, in higher than five dimensions
no wormhole exists. They also showed, for five dimensional spacetime,
wormhole exists in restricted region. On the other hand, in the usual four
dimensional spacetime, a stable wormhole was obtained which is
asymptotically flat. In presence of pressure anisotropy the model of compact star was developed by several authors \cite{matondo,bhar,her1,her2,aloma} in presence of conformal symmetry. Maharaj and Maartens \cite{maha1} obtained a solution for anisotropy compact star
with uniform density. But in reality, most of the star have variable density. Inspired by this fact, Gokhroo and Mehra \cite{gm} obtained a solution for an anisotropic sphere by choosing a variable density distribution, which is maximum at the center and
decreasing along the radius. \par

In 1916, after the discovery of the static interior solution proposed by Schwarzschild, more than 100 interior solutions were attempted by several researchers.  A comprehensive lists of interior solutions was publised by Stephani
et al. \cite{7} and Delgaty and Lake \cite{9}. Out of these
solutions, only a few
satisfied elementary physical requirements and could be
used to model relativistic stars.
A natural question concerned to spherically symmetric
relativistic static objects is to determine an upper bound on the compactness ratio $m/r$, where $m$ is the ADM mass
and $r$ is the radius of the boundary of the static object. Buchdahl \cite{buch} shows that a spherically symmetric isotropic
object for which the energy density is non-increasing outwards satisfies the bound $m/r < 4/9$. Since this bound was
obtained for a class of isotropic fluid sphere violating the dominant energy condition, several researchers have worked
in this area to find a sharp bound of maximum allowable ratio of mass to the radius for anisotropic spheres satisfying
all the energy conditions. Sharp bounds on $2m/r$ for static spherical objects under a variety of assumptions on the
eigenvalues of the Einstein tensor was obtained by Karageorgis and Stalker \cite{karag}. Bounds on $m/r$ for static objects with
a positive cosmological constant was obtained by Andr\'{e}asson and B\"{o}hmer \cite{boh}. Ivanov \cite {iva18} obtained a physically realistic stellar model by taking a simple
expression for the energy density and conformally flat spacetime. All the physically acceptable conditions are discussed by the
author without graphic proofs. Compact star model in modified gravity has also been obtained by several authors \cite{v1,v2,v3,v4,v5}.
 \par
 In recent past many researchers used the Karmakar \cite{kar} condition to obtain the new model of the anisotropic compact stars. Under Karmakar condition the two metric potentials of the underlying spacetime are connected by a bridge equation, i.e., $e^{\nu}$ and $e^{\lambda}$ are interconnected. In this case if one fixed one metric potential then the other metric potential can be suitable obtained and in this case the model of the compact star can be obtained by a proper choice of $e^{\mu}$ or $e^{\lambda}$. Very recently Bhar \cite{pb10} proposed a model
of anisotropic compact star obeying all the necessary physical
requirements which have been analyzed with the help
of the graphical representation. The metric potential $e^{\lambda}$ depends on $n$ and the model has been analyzed for a wide range of $n$ ($-200\leq n \leq $200). The model of compact star both charged and uncharged are studied by several authors \cite{em1,em2,em3,em4}.\par

In our present paper we are interested to present a new model of compact star by assuming pressure anisotropy. Our paper is organized as follows: In Sect. \ref{sec2}, the field equations are given, in Sect.~\ref{sec3}, we have generated a new model.
In Sect. \ref{sec4} we have matched our interior spacetime to the exterior Schwarzschild vacuum solution at the boundary of the compact star and to avoid the discontinuity of the tangential pressure, the Darmois Israel junction condition is discussed as well. Section \ref{sec5} is devoted to discuss some physical properties of the model. The stability analysis and the equilibrium conditions of the model are discussed in the next two sections. The energy conditions of the model is discussed in \ref{sec8}. In next section, we have given the relation between the mass to the radius of the compact objects, whereas the generating functions of the model is obtained in sect.~\ref{sec10}. Sect.~\ref{sec11} contains some discussion and concluding remarks.

\section{Interior Spacetime and Einstein field Equations}\label{sec2}
A static and spherically symmetry spacetime in (3+1)-dimension is described by the line following element,
\begin{equation}
ds^{2}=-A^{2}dt^{2}+B^{2} dr^{2}+r^{2}(d\theta^{2}+\sin^{2}\theta d\phi^{2}),
\end{equation}
Where the metrics $A$ and $B$ are static, i.e., functions of the radial coordinate `$r$' only.

We also assume that the matter within the star is anisotropic in nature and therefore, we write the corresponding energy-momentum tensor as,
\begin{equation}
T_{\nu}^{\mu}=(\rho+p_r)\chi^{\mu}\chi_{\nu}-p_t g_{\nu}^{\mu}+(p_r-p_t)\eta^{\mu}\eta_{\nu}
\end{equation}
with $ \chi^{i}\chi_{j} =-\eta^{i}\eta_j = 1 $ and $\chi^{i}\eta_j= 0$. Here the vector $\eta^{i}$ is the space-like vector and $\chi_i$ is the fluid 4-velocity and which is orthogonal to $ \eta^{i}$, $\rho$ is the matter density, $p_t$ and $p_r$ are respectively the transverse and radial pressure of the fluid and these two pressure components acts in the perpendicular direction to each other. The difference between these two pressures, i.e., $p_t-p_r$ is called the anisotropic factor and it is denoted by $\Delta$. This anisotropic factor measures the anisotropy inside the stellar interior and it creates an anisotropic force which is defined as $\frac{2\Delta}{r}$. This force may be positive or negative by depending on the sign of $\Delta$ but at the center of the star the force is zero since the anisotropic factor vanishes there.\par

The Einstein field equations assuming $G=1=c$ are given by
\begin{eqnarray}
8\pi \rho&=&\frac{1}{r^2}\left(1-\frac{1}{B^2}\right)+\frac{2B'}{B^3r},\label{1a}\\
8\pi p_r&=&\frac{1}{B^2}\left(\frac{1}{r^2}+\frac{2A'}{Ar}\right)-\frac{1}{r^2},\label{2a}\\
8\pi p_t&=&\frac{A''}{AB^2}-\frac{A'B'}{AB^3}+\frac{1}{B^3rA}(A'B-B'A).\label{3a}
\end{eqnarray}

Where `prime' indicates differentiation with respect to radial co-ordinate $r$. \\
The mass function, $m(r)$, within the radius `$r$' is given by,
\begin{equation}
m(r)=4\pi\int_0^{r}\omega^{2}\rho(\omega)d\omega.\label{4}
\end{equation}

\begin{figure}[htbp]
    \centering
        \includegraphics[scale=0.9]{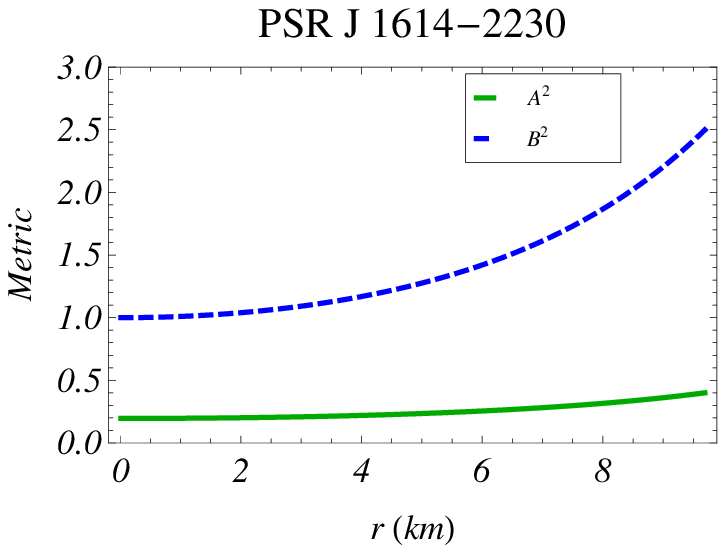}
        \includegraphics[scale=0.9]{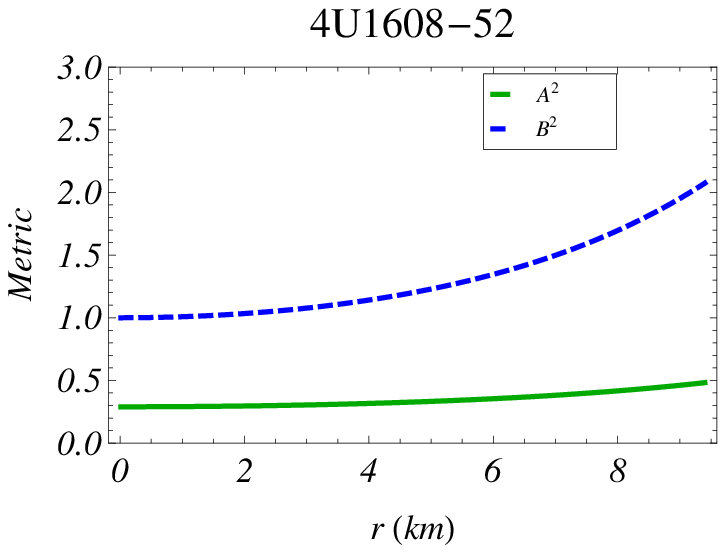}
       \caption{The metric potentials are plotted against $r$ inside the stellar interior for the compact star PSR J 1614-2230 (top panel) and 4U1608-52 (bottom panel) by taking the values of the constants $a,\,C$ and $D$ mentioned in table~1.}
    \label{metric}
\end{figure}

\section{The new anisotropic solution}\label{sec3}
The role of pressure anisotropy in modeling compact objects in the context of general theory of relativity
has been discussed by several authors \cite{17,ha,ha1,ha2}.
Our purpose here
is to generate an exact solution which does not suffer from singularities.
To solve the above set of equations (\ref{1a})-(\ref{3a}), let us take the metric potential $g_{rr}$ as proposed by \cite{kb} and is given by,
\begin{equation}\label{5}
B^2=e^{ar^2},
\end{equation}
Where `$a$' is a constant which can be obtained from the matching condition.
The KB metric is an very interesting platform to construct
the compact star since it does not allow any geometrical singularity. Several investigations using the KB metric as a
seed solution can be found in the literature to model charged as well as uncharged model of compact objects.\par
Using the expression of $B^2$ from (\ref{5}) into (\ref{1a}), we obtain the matter density of the star as,
\begin{equation}\label{7}
8\pi\rho=\frac{(1 + e^{-a r^2}) (-1 + 2 a r^2)}{r^2}.
\end{equation}

\begin{figure}[htbp]
    \centering
        \includegraphics[scale=.7]{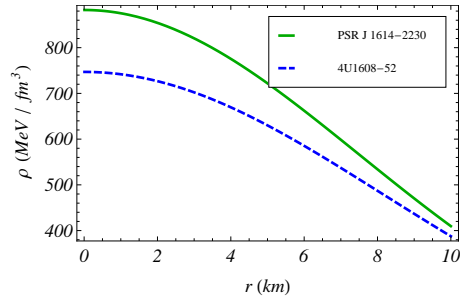}
       \caption{The matter density $\rho$ is plotted against $r$ inside the stellar interior for the compact star PSR J 1614-2230 and 4U1608-52 by taking the values of the constants $a,\,C$ and $D$ mentioned in table~1.}
    \label{rho}
\end{figure}

\begin{figure}[htbp]
    \centering
        \includegraphics[scale=.7]{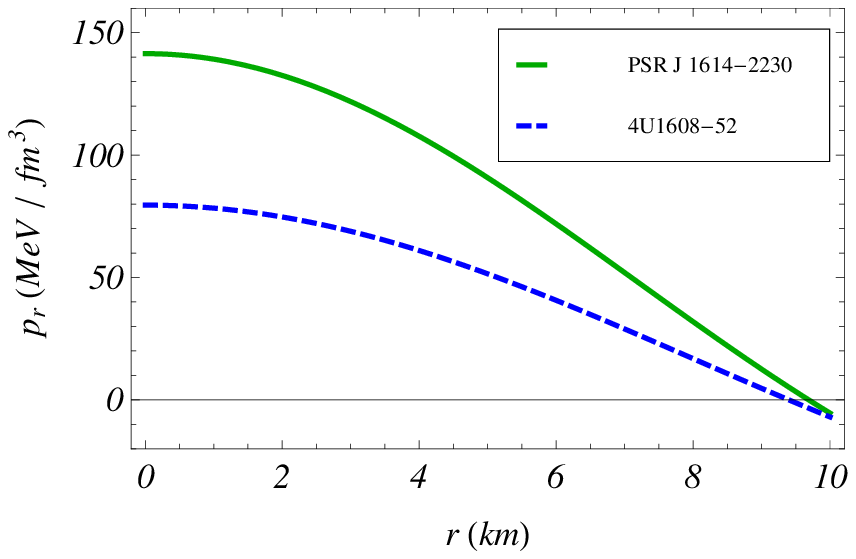}
         \includegraphics[scale=.7]{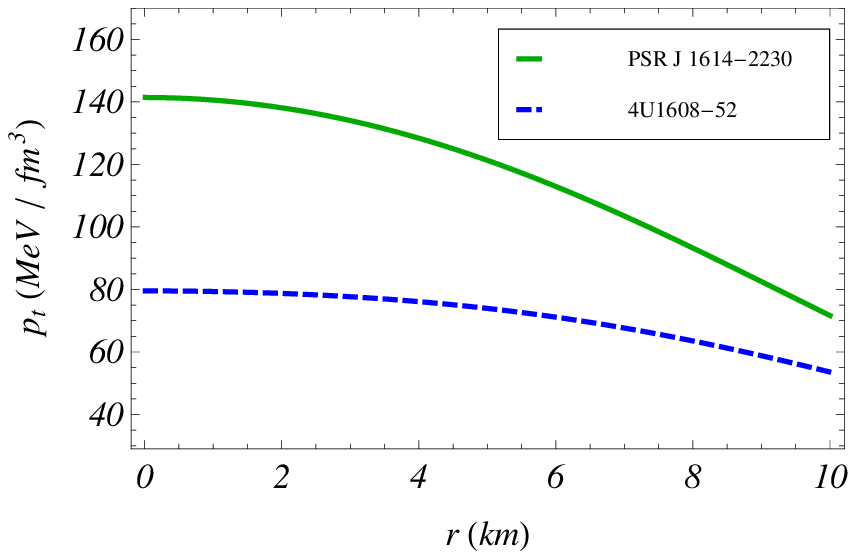}
       \caption{The radial pressure $p_r$ (top panel) and transverse pressure $p_t$ (bottom panel) are plotted against $r$ inside the stellar interior for the compact star PSR J 1614-2230 and 4U1608-52 by taking the values of the constants $a,\,C$ and $D$ mentioned in table~1.}
    \label{pr}
\end{figure}

Now using the expression of $B^2$, the expressions for radial and transverse pressure from eqns. (\ref{2a}) and (\ref{3a}) takes the form,
\begin{eqnarray}
8\pi p_r&=&e^{-ar^2}\left(\frac{1}{r^2}+\frac{2A'}{Ar}\right)-\frac{1}{r^2},\label{8}\\
8\pi p_t&=&\frac{A''}{A}e^{-ar^2}-\frac{A'}{A}a e^{-a r^2}r+\frac{e^{-ar^2}}{rA}(A'-arA).\label{9}\nonumber\\
\end{eqnarray}
Using eqns. (\ref{8}) and (\ref{9}), and introducing the anisotropic factor $\Delta=p_t-p_r$, we get the following equation:
\begin{eqnarray}\label{10}
\frac{A''}{A}-\frac{A'}{A}\frac{1 + a r^2}{r}-a - \frac{(1 - e^{a r^2})}{r^2}=\kappa e^{ar^2}\Delta,
\end{eqnarray}
Where $\kappa=8\pi$. Now our aim is to find the expression for the metric potential $A$. For this purpose, we want to solve the eqn. (\ref{10}). To integrate the equation (\ref{10}) let us take the anisotropic factor $\Delta$ in the form
\begin{equation}\label{11}
\kappa \Delta =\frac{1 - e^{-a r^2} (1 + a r^2)}{r^2}.
\end{equation}

\begin{figure}[htbp]
    \centering
        \includegraphics[scale=.9]{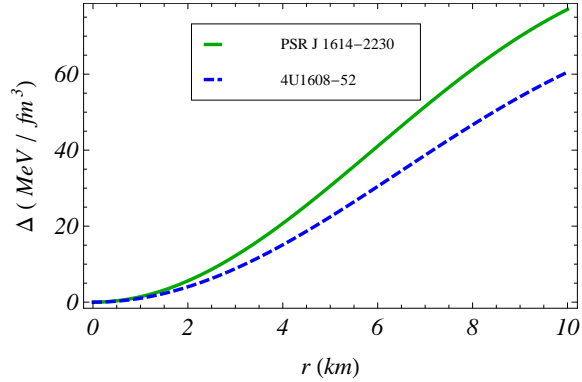}
       \caption{The anisotropic factor $\Delta$ is plotted against $r$ inside the stellar interior for the compact star PSR J 1614-2230 and 4U1608-52 by taking the values of the constants $a,\,C$ and $D$ mentioned in table~1.}
    \label{delta}
\end{figure}

If we expand $e^{-ar^2}$ in Taylor series expansion in the neighborhood of $r=0$, we get

\begin{eqnarray}\label{12}
  e^{-ar^2} &=& 1-ar^2+\frac{a^2r^4}{2!}-\frac{a^3r^6}{3!}+... ,
\end{eqnarray}
Employing the expression of (\ref{12}) into (\ref{11}), one can easily check that $\lim_{r\rightarrow0}~\Delta=0$. Moreover it provides a positive anisotropic factor inside the stellar interior which will be discussed in details in the coming sections.
With the help of this anisotropic factor equation (\ref{10}) gives,
\begin{eqnarray}
A''=A'\frac{1 + a r^2}{r}.
\end{eqnarray}
Solving the above equation, we obtain the expression of the metric coefficient for $A$ as,
\begin{eqnarray}
  A^2 &=& \left(D + \frac{C}{a} e^{\frac{a r^2}{2}}\right)^2.
\end{eqnarray}
Where $C,\,D$ are constants of integration, which can be obtained from the boundary conditions.

The radial and transverse pressure can be obtained as,
\begin{eqnarray}
  \kappa p_r &=& \frac{2 a C}{a D e^{\frac{ar^2}{2}}+ C e^{a r^2}}-\frac{1-e^{-a r^2}}{r^2}, \\
 \kappa p_t &=&\frac{a \left(-a D + C e^{\frac{a r^2}{2}}\right)}{a D e^{a r^2} + C e^{\frac{3 a r^2}{2}}}.
\end{eqnarray}
The anisotropic factor $\Delta=p_t-p_r$ is given in (\ref{11}).

\begin{figure}[htbp]
    \centering
        \includegraphics[scale=.7]{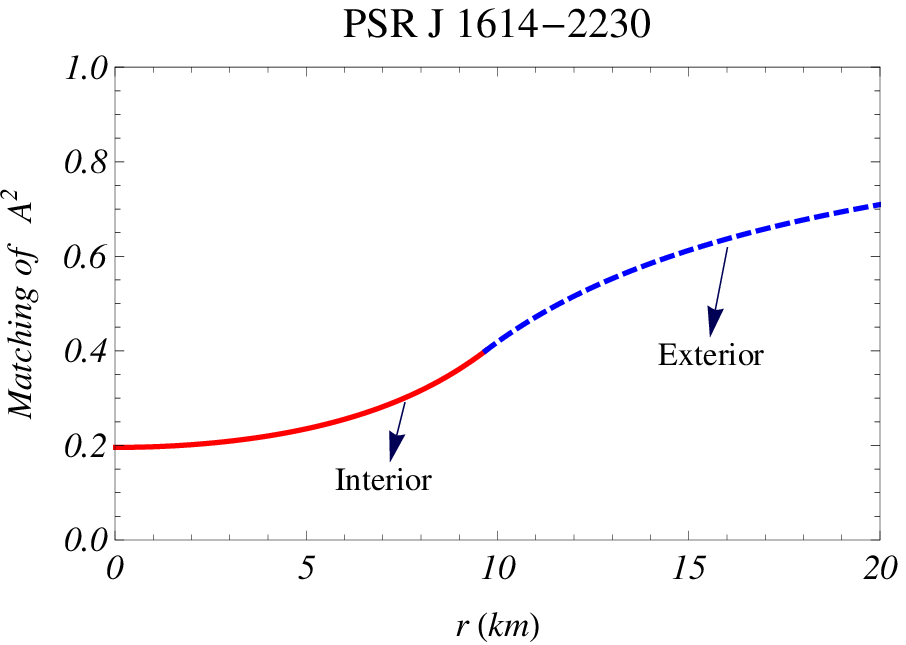}
        \includegraphics[scale=.7]{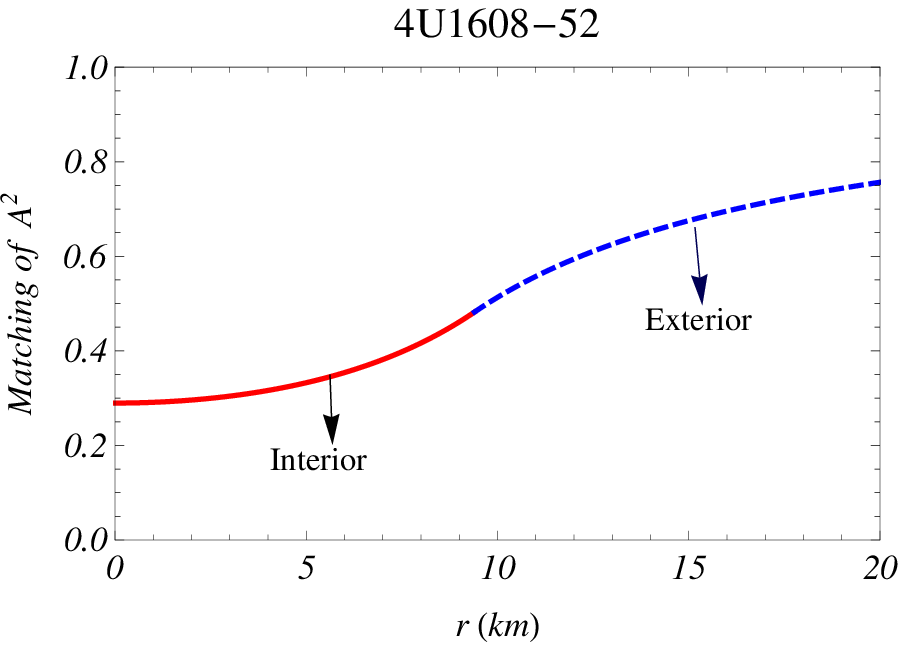}
       \caption{The matching condition of the metric potential $A^2$ is shown against $r$ for the compact star PSR J 1614-2230 (top panel) and 4U1608-52 (bottom panel) by taking the values of the constants $a,\,C$ and $D$ mentioned in table~1.}
    \label{bou1}
\end{figure}

\begin{figure}[htbp]
    \centering
          \includegraphics[scale=.7]{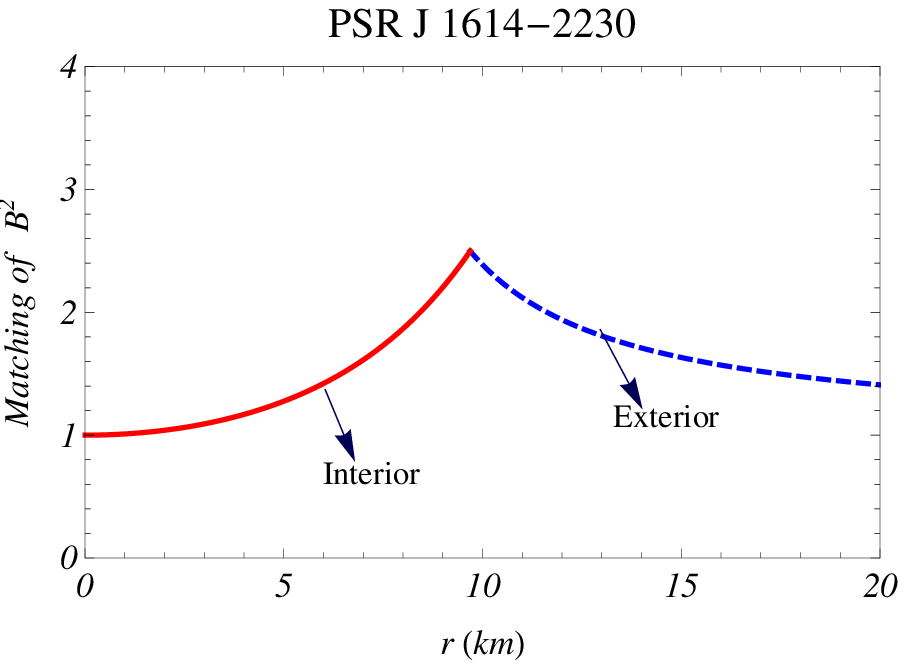}
            \includegraphics[scale=.7]{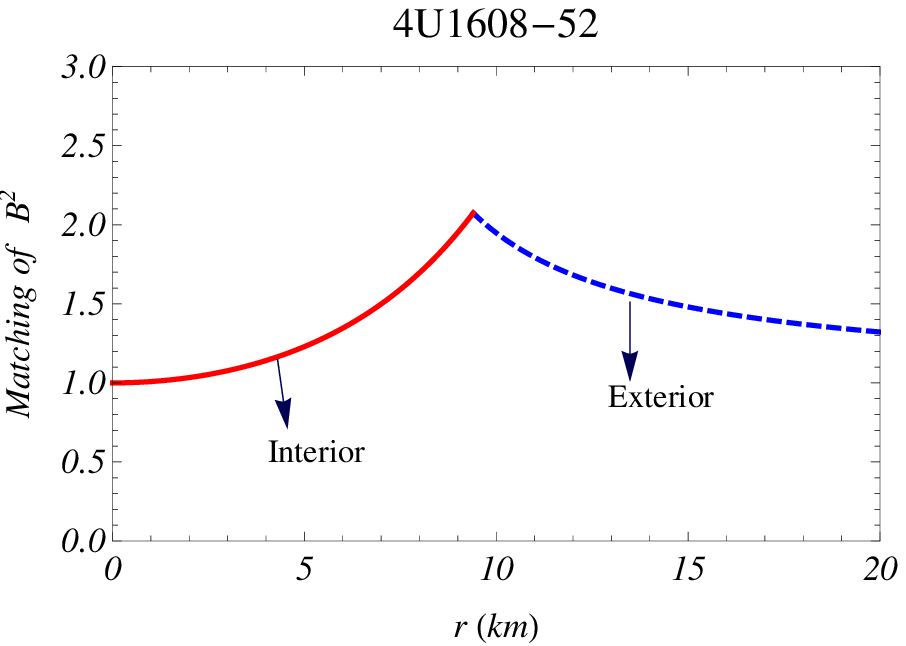}
       \caption{The matching condition of the metric potential $B^2$ is shown against $r$ for the compact star PSR J 1614-2230 (top panel) and 4U1608-52 (bottom panel) by taking the values of the constants $a,\,C$ and $D$ mentioned in table~1.}
    \label{bou2}
\end{figure}

\section{Boundary conditions}\label{sec4}
To find the constants we match our interior solution to the exterior solution smoothly at the boundary $r=R$. It is well known that the Schwarzschild vacuum solution matches exactly with the interior solution at the boundary of the star. The exterior metric is given by,
\begin{eqnarray}
ds_{+}^2 & = & -\left(1-{2m \over r}\right)dt^2+\left(1-{2m \over r}\right)^{-1}dr^2+ r^2(d\theta^2+\sin^2 \theta d\phi^2),
\end{eqnarray}
corresponding to our interior line element,
 \begin{eqnarray}
ds_{-}^2 & = & - \left(D + \frac{C}{a} e^{\frac{a r^2}{2}}\right)^2 dt^2+e^{ar^2} dr^2+ r^2(d\theta^2+\sin^2 \theta d\phi^2),
\end{eqnarray}
The first fundamental
form provides a smooth matching of the metric potentials across the boundary, i.e., at the boundary $r=R$,
\begin{eqnarray}\label{b1}
g_{rr}^+=g_{rr}^-,~g_{tt}^+=g_{tt}^-,
\end{eqnarray}
and the second fundamental form implies
\begin{eqnarray}\label{b2}
p_i(r=R-0)=p_i(r=R+0).
\end{eqnarray}
where $i$ takes the value $r$ and $t$.\\
Equations (\ref{b1})gives,
\begin{eqnarray}
  \left(1-{2M \over R}\right)= \left(D + \frac{C}{a} e^{\frac{a R^2}{2}}\right)^2, \label{k1} \\
  \left(1-{2M \over R}\right)^{-1}= e^{aR^2},\label{k2}
\end{eqnarray}
where $M=m(R)$ and from eqn.(\ref{b2}), for $i=r$, we get,
\begin{eqnarray}\label{k3}
\frac{2 a C}{a D e^{\frac{aR^2}{2}}+ C e^{a R^2}}-\frac{1-e^{-a R^2}}{R^2}= 0.
\end{eqnarray}
Solving the three equations (\ref{k1})-(\ref{k3}) we obtain,
\begin{eqnarray}
a&=&-\frac{1}{R^2}\ln\left(1-\frac{2M}{R}\right),\\
C&=&\frac{M}{R^3},\\
D&=&\sqrt{1-\frac{2M}{R}}+\frac{M}{R}\frac{1}{\sqrt{1-\frac{2M}{R}}}\frac{1}{\ln\left(1-\frac{2M}{R}\right)}.
\end{eqnarray}
The values of $a,\,C$ and $D$ for different compact stars are obtained in table~1, but there is a discontinuity for $i=t$ from eqn.(\ref{b2}) since the tangential pressure does not vanish at the boundary of the star. To avoid
the discontinuity we calculate the surface stresses at the junction
boundary $r=R$ by using the Darmois-Israel \cite{44,45} formation. The expression for surface energy density $\sigma$ and the surface
pressure $\mathcal{P}$ at the junction surface $r = R$ are obtained
as,
\begin{eqnarray*}
  \sigma &=& -\frac{1}{4\pi R}\Big[\sqrt{1-\frac{2M}{R}}-e^{-\frac{aR^2}{2}}\Big], \\
  \mathcal{P}&=& \frac{1}{8\pi R}\left[\frac{1-\frac{M}{R}}{\sqrt{1-\frac{2M}{R}}}-e^{-\frac{aR^2}{2}}-\frac{aCR^2}{aD+Ce^\frac{aR^2}{2}}\right].
\end{eqnarray*}

\begin{table}[ht]
\tbl{The values of the constants $a,\,C$ and $D$ of few well known compact star candidates.}
{\begin{tabular}{@{}ccccccccc@{}}\toprule
Objects & Observed & Observed  & Estimated & Estimated & $a$ & $C$  &$D$ & References\\
& mass ($M_{\odot}$)& radius & Mass ($M_{\odot}$)& Radius &(km$^{-2}$)&(km$^{-2}$)\\  \colrule
Vela X -1& $1.77 \pm 0.08$ & $9.56 \pm 0.08$&$1.77$&$9.5$&$0.00883866$&$0.00304505$&$0.157733$& \cite{raw}\\
LMC X -4 & $1.04 \pm 0.09$&$8.301 \pm 0.2$ &$1.05$&$8.1$&$ 0.00734532$&$0.00291425$&$0.281018$  &\cite{raw}             \\
4U 1608 - 52   &$1.74 \pm 0.14$ &$9.528\pm 0.15$  &  $1.65$&$9.4$&$ 0.00825527$&$0.00293017$&$0.183234$ &\cite{guv}  \\
PSR J1614 - 2230&$1.97 \pm 0.04$ & $9.69 \pm 0.2$&$ 1.97$&$9.69$&$0.00975169$&$0.00319365$&$0.115009$ &\cite{demo}      \\
EXO 1785 - 248&$1.3 \pm 0.2$  &$8.849 \pm 0.4$   &$1.4$&$9$&  $0.00758186$&$0.00283265$&$0.227708$&\cite{ozel}            \\   \botrule
\end{tabular}}
\end{table}

\section{Physical attributes}\label{sec5}
\begin{itemize}
\item  {\bf Regularity of the metric coefficients :} For a physically acceptable model, solution should be free from physical and geometric singularities, i.e., it should take finite and positive values for the the metric potentials. Now $A|_{r=0}=D+\frac{C}{a}>0$ and  $B|_{r=0}=1$. We have drawn the profiles of the metric co-efficients for the compact star against $r$ in Fig~\ref{metric} for the compact stars PSR J 1614-2230 and 4U1608-52 respectively. We have also matched the interior metric co-efficients to the exterior spacetime in figs.~\ref{bou1},\ref{bou2}.\\
\begin{figure}[htbp]
    \centering
        \includegraphics[scale=.8]{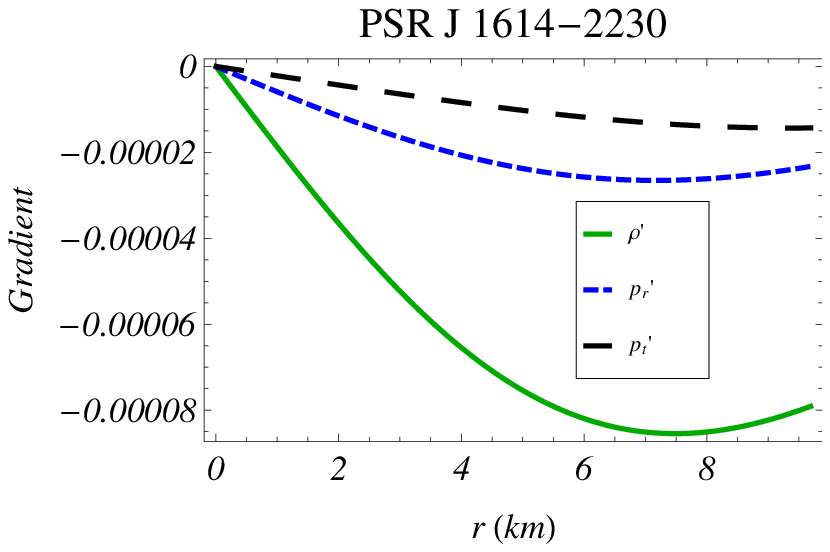}
         \includegraphics[scale=.8]{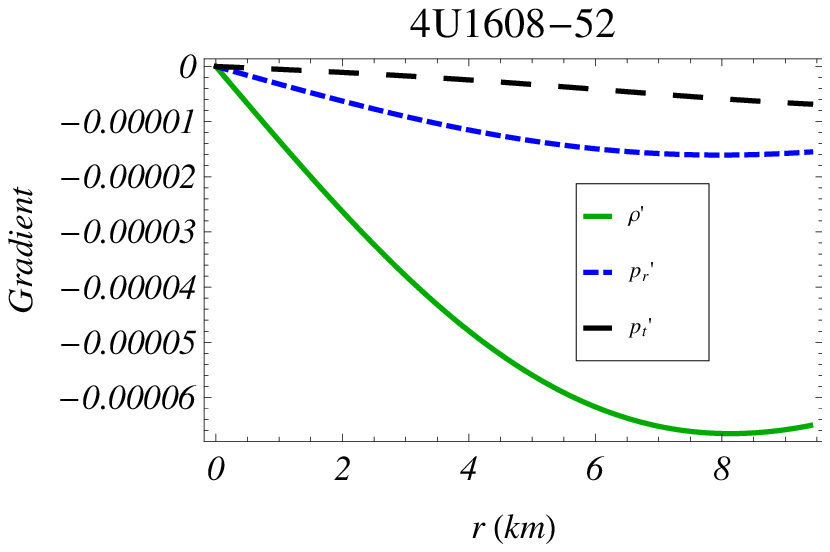}
       \caption{The pressure and density gradients are plotted against $r$ inside the stellar interior for the compact star PSR J 1614-2230 (top panel) and 4U1608-52 (bottom panel) by taking the values of the constants $a,\,C$ and $D$ mentioned in table~1.}
    \label{grad}
\end{figure}
    \item {\bf Regularity of the density and pressure :} The central pressure, central density should be nonzero positive value inside the stellar interior, where as $p_r$ should vanish at the boundary. The central pressures and the density can be written as
\begin{eqnarray}
p_{r0}= p_{t0}=p_c=\frac{1}{\kappa}\frac{a (C - a D)}{C + a D}>0 \\
\rho_c=\frac {3a}{\kappa}>0
\end{eqnarray}
The above two inequalities provide the following restrictions on the parameters:
\begin{eqnarray}\label{in}
a>0,~C>aD,
\end{eqnarray}
The surface density of the compact star model is obtained as,
\begin{eqnarray}
\rho_s&=&\frac{e^{-a R^2}\big(e^{a R^2}+ 2 a R^2-1\big)}{\kappa R^2}.
\end{eqnarray}
The numerical values of the central density and surface density for different compact star model is obtained in table~2.
To find the behavior of the matter density ($\rho$) and radial and transverse pressure $p_r$ and $p_t$ inside the stellar interior, the profiles of $\rho$ and $p_r,\,p_t$ are shown in Fig.~\ref{rho} and Fig.~\ref{pr} respectively.\\
The density and pressure gradient of our present model is obtained as,
\begin{eqnarray}
  \kappa \rho' &=& \frac{2 \big\{-1 + e^{-a r^2} (1 + a r^2 - 2 a^2 r^4)\big\}}{r^3}, \\
 \kappa p_r' &=& \frac{2 - 2 e^{-a r^2}}{r^3} - \frac{2 a e^{-a r^2}}{r} - \frac{
 2 a^2 C e^{-\frac{a r^2}{2}} \left(a D + 2 C e^{-\frac{a r^2}{2}}\right) r}{\left(a D +
   C e^{-\frac{a r^2}{2}}\right)^2},\\
   \kappa p_t'&=&\frac{2 a^2 e^{-a r^2}\left(a^2 D^2 + a C D e^{-\frac{a r^2}{2}} -
   C^2 e^{a r^2}\right) r}{\left(a D + C e^{-\frac{a r^2}{2}}\right)^2}.\nonumber\\
\end{eqnarray}
Now at the point $r=0$,
$\rho'=0,\,p_r'=0,\,p_t'=0$ also,
\begin{eqnarray}
  \kappa\rho'' &=&-5a^2<0,  \\
  \kappa p_r''&=& \frac{-3 a^2 C^2 + a^4 D^2}{(C + a D)^2}, \\
  \kappa p_t''&=& \frac{2 a^2 (-C^2 + a C D + a^2 D^2)}{(C + a D)^2}.
\end{eqnarray}
Both $p_r'$ and $p_t'$ are negative since from (\ref{in}) we get $C>aD$. Which indicates that $\rho,\,p_r$ and $p_t$ all are monotonic decreasing function of $r$, they take maximum value at the center of the star and then gradually decreases towards the boundary. Also both the density and pressure gradients are negative (Fig.\ref{grad}), it is once again verified that both density and pressures are monotonic decreasing function of $r$.\\
\item {\bf Nature of pressure anisotropy :}
To investigate the behavior of pressure anisotropic for our present model of compact star the profile of $\Delta$ has been shown in fig.~\ref{delta}. The figure shows that the the pressure anisotropic is positive and consequently the anisotropic force is positive inside the stellar interior which creates a repulsive force towards the boundary of the star and this positive anisotropy helps balance the gravitational force acting on the stellar model and keep up the star from collapsing.
\begin{figure}[htbp]
    \centering
        \includegraphics[scale=.8]{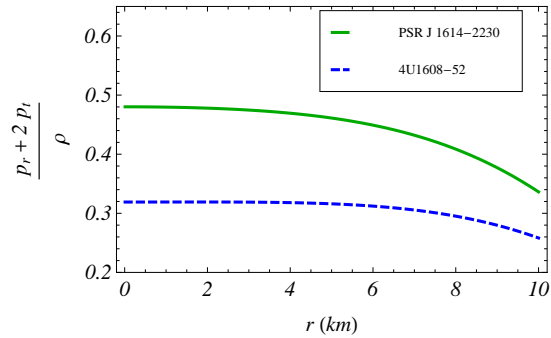}
       \caption{$\frac{p_r+2p_t}{\rho}$ is plotted against $r$ inside the stellar interior for the compact star PSR J 1614-2230 and 4U1608-52 by taking the values of the constants $a,\,C$ and $D$ mentioned in table~1.}
    \label{diag}
\end{figure}
\item Bondi \cite{bondi} proposed that, for an anisotropic fluid sphere $(p_r+2p_t)/\rho$ should be less than $1$. We have plotted  $(p_r+2p_t)/\rho$ vs $r$ in Fig.~\ref{diag} to check this condition for the compact star PSR J 1614-2230 and 4U1608-52. It is clear from the figure that our model satisfies the condition of Bondi.
\begin{figure}[htbp]
    \centering
        \includegraphics[scale=.8]{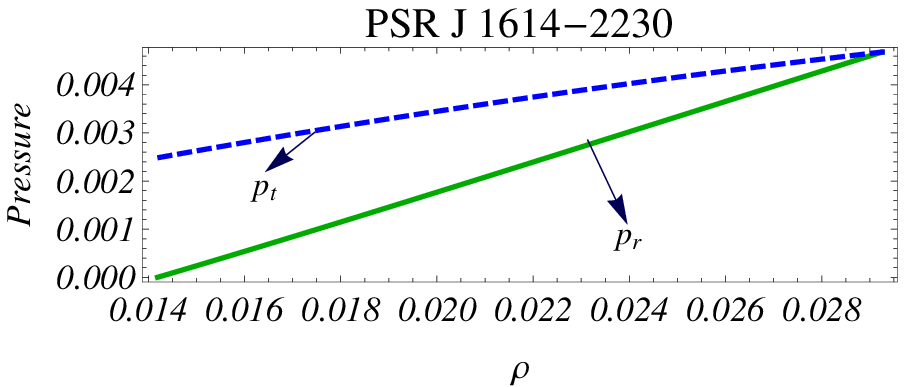}
         \includegraphics[scale=.8]{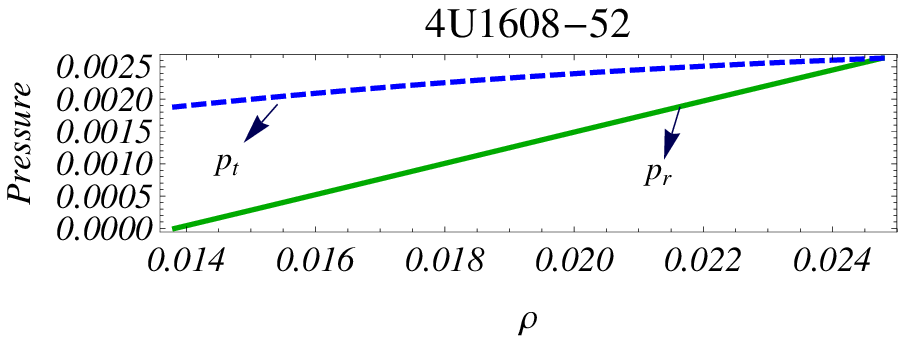}
       \caption{The relation between the density $\rho$ and pressure (both $p_r$ and $p_t$) are plotted inside the stellar interior for the compact star PSR J 1614-2230 (top panel) and 4U1608-52 (bottom panel) by taking the values of the constants $a,\,C$ and $D$ mentioned in table~1.}
    \label{eos}
\end{figure}
    \item {\bf Equation of state} In order to develop an anisotropic stellar model by solving the Einstein's field equation, it is a common practice to consider
a relationship between matter variables of the fluid
configuration with the pressure $p_r$ and $p_t$ which is usually known as equation of state (EoS). Many researchers used linear or non linear or nonlinear EoS to develop the model of the compact star. Among the linear EoS, MIT bag model EoS for quark
matter defined by, $p_r=\frac{1}{3}(\rho-4B_g)$ is a very interest choice to the researchers, where $B_g$ is the bag constant \cite{d1,d2,d3,d4,d5}. To develop our present model of the compact star, instead of choosing any EoS, we have chosen a physically reasonable anisotropic factor $\Delta$. For this reason it is not possible to find an analytical relation between the density to the pressure, i.e., why we have taken the help of the graphical representation. The relation between the pressure and density have been shown in fig.~\ref{eos}. It is clear from the figure that, both radial and transverse pressure maintain almost a linear relationship with the matter density.

\end{itemize}

\begin{table}[t]
\tbl{The numerical values of the central density ($\rho_c$), surface density ($\rho_s$), central pressure $p_c$ and surface redshift of few well known compact star candidates.}
{\begin{tabular}{@{}cccccccccc@{}}\toprule
Objects & $\rho_c(gm.cm^{-3})$&$\rho_s(gm.cm^{-3})$&$p_c(dyne.cm^{-2})$&$\frac{2M}{R}$&$Z_s$\\  \colrule
Vela X -1& $1.42359\times10^{15}$&$7.54391\times10^{14}$&$1.58827\times10^{35}$&$0.549632$&$0.490102$\\
LMC X -4  &$1.18307\times10^{15}$&$8.00021\times10^{14}$ &$0.60604\times10^{35}$&$0.382407$ &$0.272474 $  \\
4U 1608 - 52 &$1.32963\times10^{15}$&$7.42043\times10^{14}$&$1.27269\times10^{35}$ &   $0.517819$&$0.440108$  \\
PSR J1614 - 2230  &$1.57065\times10^{15}$ &$7.6203\times10^{14}$  &$2.26264\times10^{35}$ &$0.599742$&$0.580629$     \\
EXO 1785 - 248&$1.22116\times10^{15}$&$7.44681\times10^{14}$&$0.88889\times10^{35}$  &$0.458889$&$0.359430  $      \\  \botrule
\end{tabular}}
\end{table}

\section{Stability condition}\label{sec6}
In this section we want to discuss the stability of the present model via (i) Harrison-Zeldovich-Novikov condition, (ii) Causality Condition and Herrera's method of cracking and (iii) Relativistic adiabatic index.
\subsection{Stability due to Harrison-Zeldovich-Novikov }
Harrison et al. \cite{harri} and Zeldovich-Novikov \cite{novi} proposed a stability condition for the model of compact star which depends on the mass and central density. They proved that a stellar configuration will be stable if $\frac{\partial M}{\partial \rho_c}>0$. Where $M,\,\rho_c$ denotes the mass and central density of the compact star.
\begin{figure}[htbp]
    \centering
        \includegraphics[scale=.8]{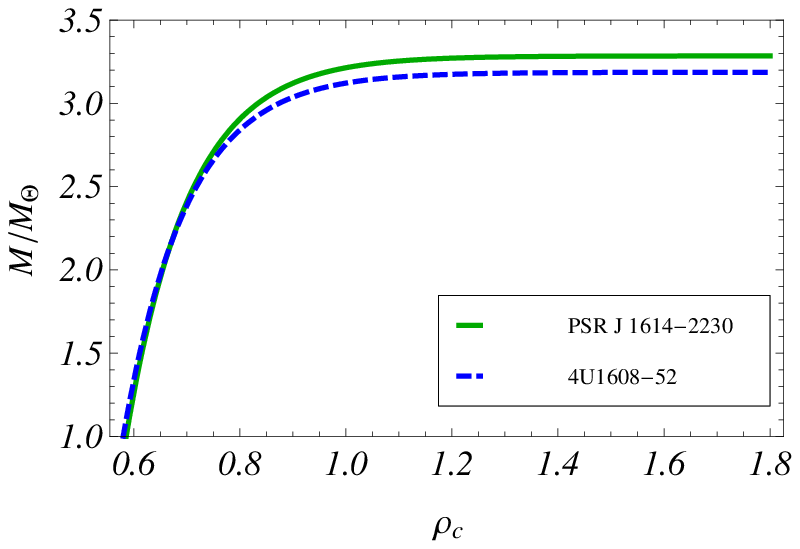}
         \includegraphics[scale=.8]{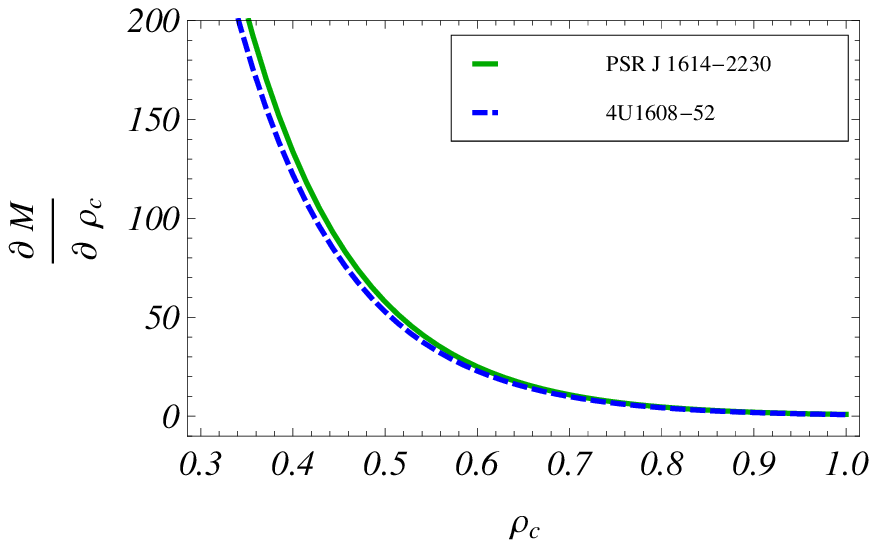}
       \caption{(top panel) The variation of the mass function and (bottom panel) $\frac{\partial M}{\partial \rho_c}$ are plotted with respect to central density $\rho_c$ inside the stellar interior for the compact star PSR J 1614-2230 and 4U1608-52 by taking the values of the constants $a,\,C$ and $D$ mentioned in table~1.}
    \label{tt}
\end{figure}

For our present model,
\begin{eqnarray}
\frac{\partial M}{\partial \rho_c}=\frac{4}{3}\pi R^3 e^{-\frac{8\pi \rho_c}{3}}.
\end{eqnarray}
It is very clear from the above expression that $\frac{\partial M}{\partial \rho_c}>0$ and therefore the stability condition is well satisfied. In fig.~\ref{tt}, we have shown the variation of the mass function and $\frac{\partial M}{\partial \rho_c}$ with respect to the central density.

\subsection{Causality Condition and cracking}
\begin{figure}[htbp]
    \centering
    \includegraphics[scale=.8]{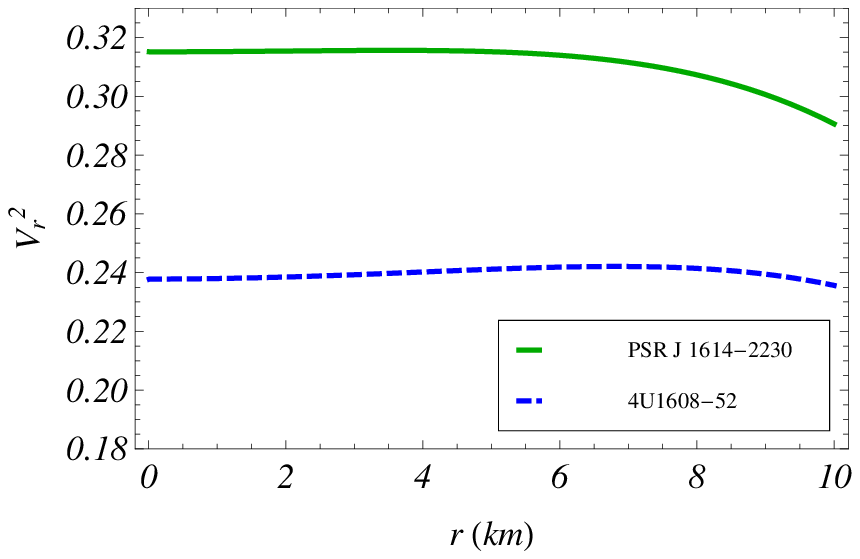}
        \includegraphics[scale=.8]{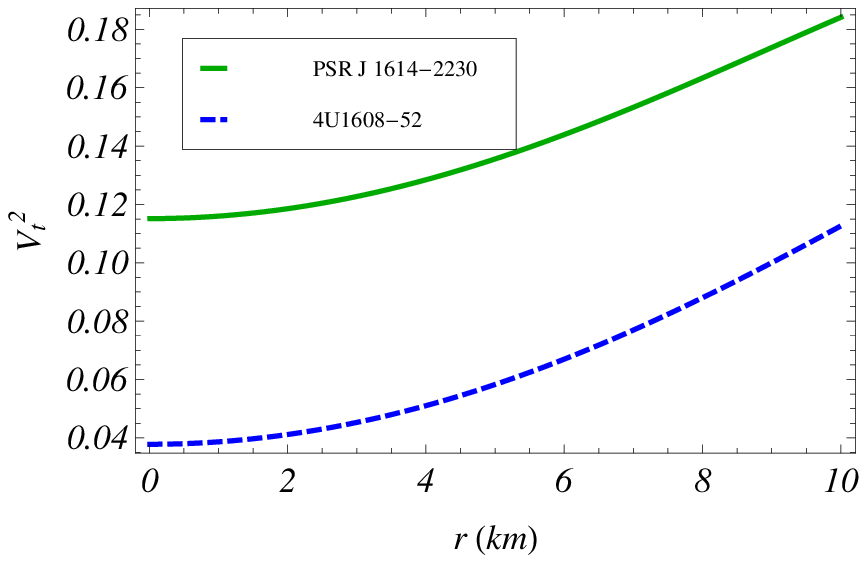}
          \includegraphics[scale=.8]{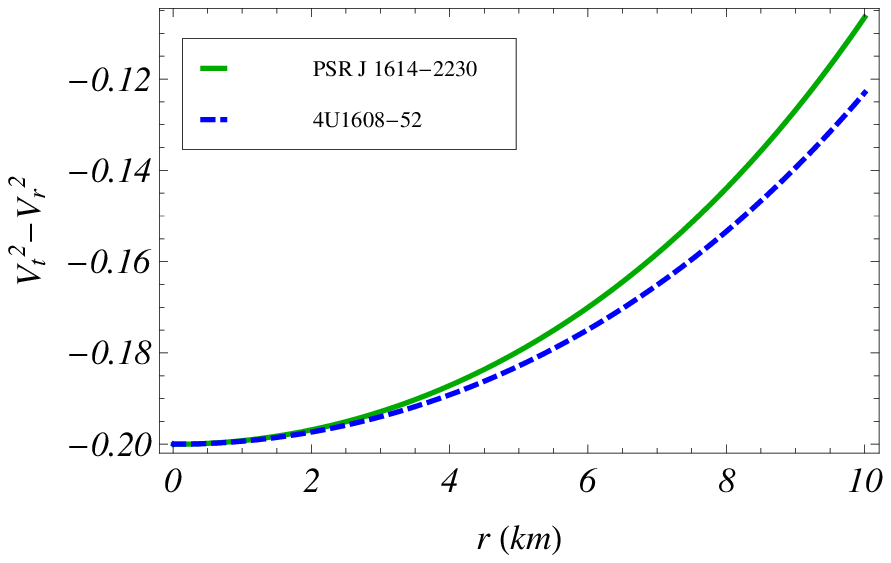}
       \caption{(Top) The variation of the square of radial velocity ($V_r^2$), (middle) the variation of the square of transverse velocity ($V_t^2$), (bottom) the variation of $V_t^2-V_r^2$ against $r$ is shown inside the stellar interior for the compact star PSR J 1614-2230 and 4U1608-52 by taking the values of the constants $a,\,C$ and $D$ mentioned in table~1.}
    \label{sv}
\end{figure}

Next we are interested to check the subliminal velocity of sound for our present model. Since we are dealing with the anisotropic fluid, the square of the radial and transverse velocity of sound $V_r^2$ and $V_t^2$ respectively should obey some bounds. According to the principle of Le Chatelier, speed of sound must be positive i.e., $V_r~>0,\,V_t>0$. At the same time, for anisotropic compact star model, both the radial and transverse velocity of sound should be less than $1$ which is known as causality conditions. Combining the above two inequalities one can obtain, $0<V_r^2,\,V_t^2<1$. For our present model,
\begin{eqnarray}
V_{r}^{2}&=&\frac{dp_r}{d\rho}\nonumber\\
&=&\frac{e^{a r^2} r^3}{2 \Big((1-a r^2)(1 + 2 a r^2)-e^{a r^2}\Big)}\times\Big[\frac{2 - 2 e^{-a r^2}}{r^3}\nonumber\\&& - \frac{2 a e^{-a r^2}}{r} -\frac{
   2 a^2 C e^{-\frac{a r^2}{2}} \big(a D + 2 C e^{\frac{a r^2}{2}} r\big)}{(a D +
     C e^{-\frac{a r^2}{2}})^2}\Big]\nonumber\\
V_{t}^{2}&=&\frac{dp_t}{d\rho}\\
&=&\frac{a^2 \Big(a^2 D^2 + a C D e^{\frac{a r^2}{2}} - C^2 e^{a r^2}\Big) r^4}{(a D +
   C e^{\frac{a r^2}{2}})^2 \Big((1-a r^2)(1 + 2 a r^2)-e^{a r^2}\Big)}\nonumber\\
\end{eqnarray}
To check the reasonable bound for $V_r^2,\,V_t^2$ , we have drawn the profiles of the squares of the radial and transverse velocity of sound in the top panel of fig.~\ref{sv} for the compact star PSR J 1614-2230 and 4U1608-52 and we note that the bound is obeyed by $V_r^2,\,V_t^2$.\par
Next we are interested to check whether our present model satisfy the ``method of cracking" proposed by Herrera \cite{her}. Using the method of `cracking', Abreu et al. \cite{abreu} proposed that, the region inside the stellar interior is potentially unstable if the square of the transverse velocity of sound is greater than or equal to the square of the radial velocity of sound otherwise the region is potentially stable. Symbolically, for Stable region $V_t^2-V_r^2<0$. Now, at the center of the star,
\[V_r^2=\frac{3 C^2 - a^2 D^2}{5 (C + a D)^2};~~V_t^2=\frac{2 (C^2 - a C D - a^2 D^2)}{5 (C + a D)^2}\]
and the difference between the squares of the velocities, i.e., $V_t^2-V_r^2=-\frac{1}{5}<0$ at the center. So, at the center, the ``cracking method'' is satisfied. However the profile of $V_t^2-V_r^2$ against $r$ inside the stellar interior for the compact star PSR J 1614-2230 and 4U1608-52 are shown in fig.~\ref{sv}(bottom panel) and the figure indicates that our proposed model of compact star is potentially stable everywhere inside the boundary.

\subsection{Relativistic Adiabatic index}
In this subsection we want to check the stability of our present model via relativistic adiabatic index. The adiabatic index $\Gamma$ is the ratio of the two specific heat and its expression can be obtained from the following formula:
\begin{eqnarray}
\Gamma&=&\frac{\rho+p_r}{p_r}V_r^2\nonumber \\
&=&\frac{2 a (a D + 2 C e^{\frac{a r^2}{2}} r^2}{a D-a D e^{a r^2}-
 Ce^{\frac{3a r^2}{2}}+C e^{\frac{a r^2}{2}} (1 + 2 a r^2)}V_r^2.\nonumber\\
\end{eqnarray}

\begin{figure}[htbp]
    \centering
        \includegraphics[scale=.8]{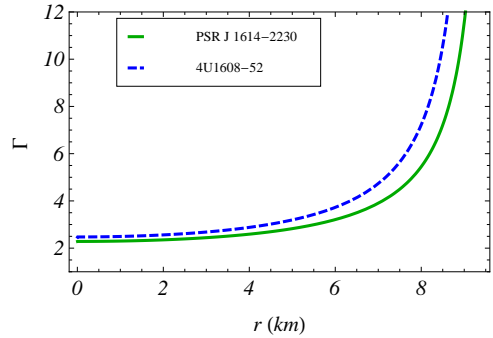}
       \caption{The adiabatic index $\Gamma$ is shown against $r$ inside the stellar interior for the compact star PSR J 1614-2230 and 4U1608-52 by taking the values of the constants $a,\,C$ and $D$ mentioned in table~1.}
    \label{gama}
\end{figure}

Now for a newtonian isotropic sphere the stability condition is given by $\Gamma>\frac{4}{3}$ and for an anisotropic collapsing stellar configuration, the condition is quite difficult and it changes to \cite{bondi64}
\begin{eqnarray}
\Gamma> {4\over 3}+\left[{4\over 3}~{p_{ti}-p_{ri} \over r|p'_{ri}|}+{8\pi r \over 3}~{\rho_i p_{ri} \over |p'_{ri}|}\right]_{max} \label{gam2}
\end{eqnarray}
here $p_{ri}, ~p_{ti}$ and $\rho_i$ are initial values of radial pressure, transverse pressure and density respectively. From eqn. (\ref{gam2}), it is clear that for a stable anisotropic configuration, the limit on adiabatic index depends upon the types of anisotropy. In our present case, we have plotted the profile of $\Gamma$ and we see that it is always greater than $\frac{4}{3}$ and hence we get stable configuration (Fig. \ref{gama}).

\section{Equilibrium condition}\label{sec7}
To check the static stability condition of our model under three different forces, the generalized Tolman-Oppenheimer-Volkov (TOV) equation has been considered which is represented by the equation
\begin{equation}\label{tov1}
-\frac{M_G(\rho+p_r)}{r^{2}}\frac{B}{A}-\frac{dp_r}{dr}+\frac{2}{r}(p_t-p_r)=0
\end{equation}
Where $M_G = M_G(r)$ is the effective gravitational mass
inside the fluid sphere of radius `$r$' and is defined by
\begin{equation}\label{tov2}
M_G(r)=r^{2}\frac{A'}{B},
\end{equation}

\begin{figure}[htbp]
    \centering
       \includegraphics[scale=.8]{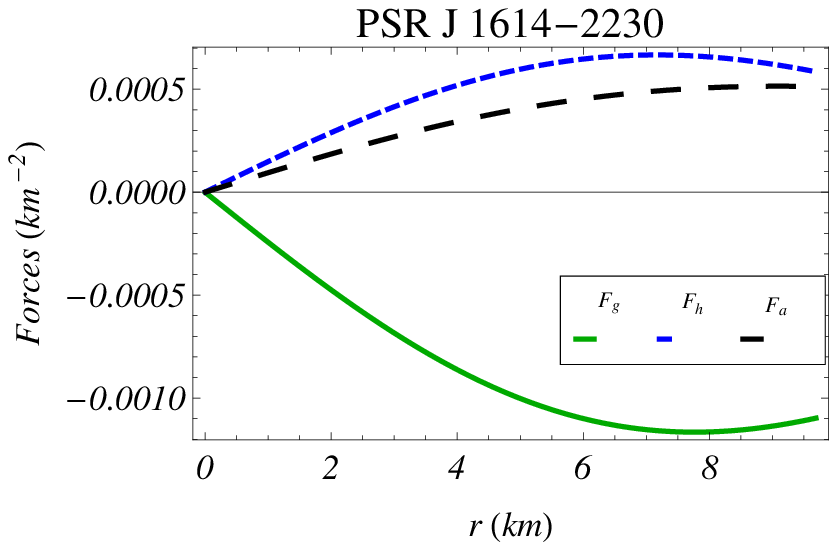}
       \includegraphics[scale=.8]{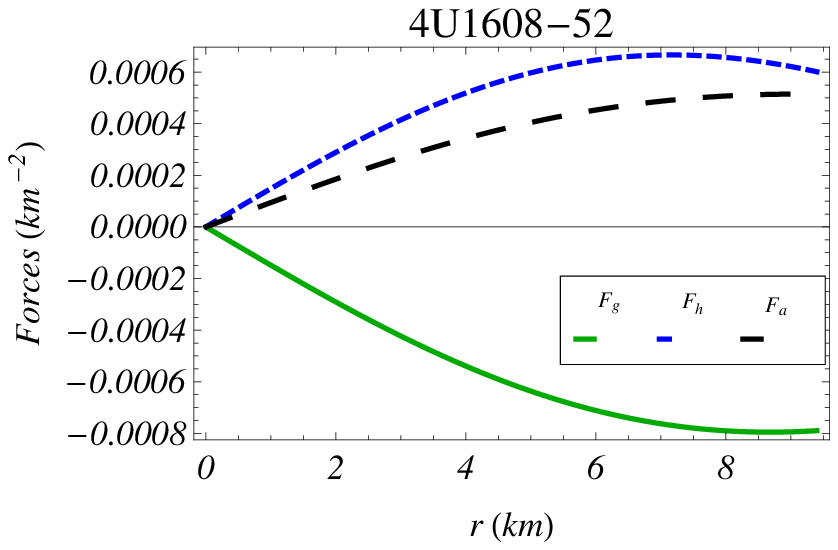}
       \caption{The variation of gravitational, hydro-statics and anisotropic forces are shown against $r$ inside the stellar interior for the compact star PSR J 1614-2230 (top panel) and 4U1608-52 (bottom panel) by taking the values of the constants $a,\,C$ and $D$ mentioned in table~1.}
    \label{tov}
\end{figure}

The above expression of $M_G(r)$ can be derived from Tolman-Whittaker mass formula. Using the expression of equation (\ref{tov2}) in (\ref{tov1}) we obtain the modified TOV equation as,
\begin{equation}
F_g+F_h+F_a=0
\end{equation}
Where the expression of the three forces are given by,
\begin{eqnarray}
F_g&=&-\frac{A'}{A}(\rho+p_r)\nonumber\\
&=&-\frac{2 a^2 C e^{-\frac{a r^2}{2}} \left(a D + 2 C e^{-\frac{a r^2}{2}}\right) r}{\kappa\left(a D +
  C e^{-\frac{a r^2}{2}}\right)^2},\\
F_h&=&-\frac{dp_r}{dr}=\frac{2}{\kappa}\Big[\frac{-1 + e^{-a r^2}}{r^3} + \frac{a e^{-a r^2}}{r}\nonumber\\&&+\frac{a^2 C e^{-\frac{a r^2}{2}}\big(a D + 2 C e^{-\frac{a r^2}{2}}\big) r}{\big(a D +C e^{-\frac{a r^2}{2}}\big)^2}\Big],\\
F_a&=&\frac{2}{r}(p_t-p_r)=\frac{2 (1 - e^{-a r^2}(1 + a r^2)}{r^3}.
\end{eqnarray}
$F_g$, $F_h$ and $F_a$ are known as gravitational, hydro-statics and anisotropic forces respectively. The profile of the above three forces for our model of compact star is shown in Fig. 12, which verifies that present system is in static equilibrium under these three forces.

\section{Energy Conditions}\label{sec8}
It is well known that for a compact star model the energy conditions should be satisfied and in this section we are interested to study about it. For an anisotropic compact star, all the energy conditions namely Weak Energy Condition (WEC), Null Energy Condition (NEC) and Strong Energy Condition (SEC)  are satisfied if and only if the following inequalities hold simultaneously for every points inside the stellar configuration.

\begin{figure}[htbp]
   \centering
        \includegraphics[scale=.8]{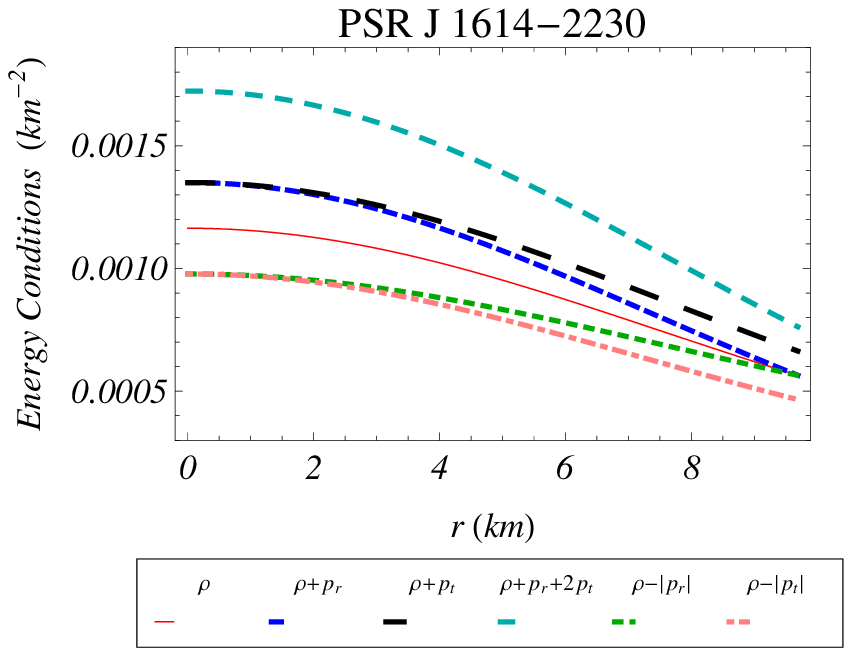}
          \includegraphics[scale=.8]{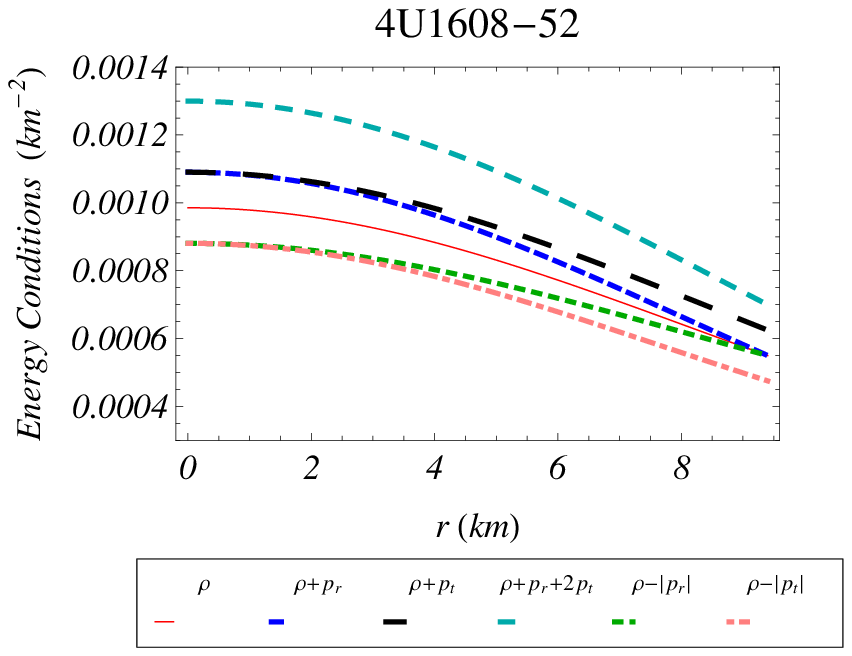}
      \caption{Energy conditions are plotted against $r$ inside the stellar interior for the compact star PSR J 1614-2230 (top panel) and 4U1608-52 (bottom panel)by taking the values of the constants $a,\,C$ and $D$ mentioned in table~1.}
  \label{ec}
\end{figure}

\begin{eqnarray}\label{1}
WEC &:& T_{\mu \nu}u^\mu u^\nu \ge 0~\mbox{or}~\rho \geq  0,~\rho+p_i \ge 0  \label{2k}\\
NEC &:& T_{\mu \nu}v^\mu v^\nu \ge 0~\mbox{or}~ \rho+p_i \geq  0\\ \label{3}
DEC &:& T_{\mu \nu}u^\mu u^\nu \ge 0 ~\mbox{or}~ \rho \ge |p_i|  \\ \label{4}
SEC &:& T_{\mu \nu}u^\mu u^\nu - {1 \over 2} T^\lambda_\lambda u^\sigma u_\sigma \ge 0 ~\mbox{or}~ \rho+\sum_i p_i \ge 0.\label{4k}
\end{eqnarray}
Where $i$ takes the value $r$ and $t$ for radial and transverse pressure. $~u^\mu$ and $v^\mu$ are time-like vector and null vector respectively and $T^{\mu \nu}u_\mu $ is nonspace-like vector. To check all the inequality stated above we have drawn the profiles of ~l.h.s of (\ref{2k})-(\ref{4k}) in fig~\ref{ec} in the interior of the compact star PSR J 1614-2230 and 4U1608-52. The figure shows that all the energy conditions are satisfied by our model.

\section{Mass-Radius relation and Redshift}\label{sec9}
Using the relationship $B^2=\left(1-\frac{2m}{r}\right)^{-1}$, the mass function of the present model can be obtained as,
\begin{equation*}\label{m1}
m(r)=\frac{r}{2}(1 - e^{-a r^2}),
\end{equation*}
We can easily check that $\lim_{r\rightarrow0}m(r)=0.$ Since $e^{ar^2}>1$ for $r>0,$ and for all $a>0$, so $e^{-ar^2}<1$ and consequently $1-e^{-ar^2}>0$. So it is clear that $m(r)>0$ for $0<r\leq~R$, R being the radius of the star. Hence, it is proved that the mass function is positive inside the stellar interior. Moreover $e^{ar^2}$ is monotonic increasing function of $r$ for $a>0$ and therefore $e^{-ar^2}$ is monotonic decreasing function of $r$ and it corresponds that $1 - e^{-a r^2}$ takes higher value for increasing $r$. It analytically verifies that the mass function is monotonic increasing with respect to radius $r$.\par
The compactness factor, i.e., the ratio of mass to the radius of a compact star for our present model is obtained from the formula,
\begin{equation}
u=\frac{m(r)}{r}=\frac{1}{2}(1 - e^{-a r^2}),
\end{equation}
\begin{figure}[htbp]
   \centering
        \includegraphics[scale=.7]{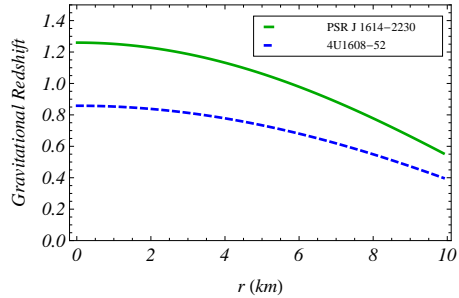}
      \caption{Gravitational redshift function is plotted against $r$ inside the stellar interior for the compact star PSR J 1614-2230 and 4U1608-52 by taking the values of the constants $a,\,C$ and $D$ mentioned in table~1.}
  \label{grav}
\end{figure}

We have already discussed that $1 - e^{-a r^2}$ is monotonic increasing function of $r$, so the maximum value of $u$ attains at the radius of the star. The maximum possible ratio of twice mass to the radius i.e., $\frac{2M}{R}$ for different compact stars are shown in Table~2. Buchdahl \cite{buch} proposed that for a compact star model, $\frac{2M}{R}$ should be less than $\frac{8}{9}$. Table~2 indicates that the values of compactness factor for different compact stars lies in the proposed range. The ratio of mass to the radius plays an important role to recognize a compact objects. The compactness factor classifies the compact star as follows \cite{kanti} : (i)  for normal star : $\frac{M}{R}\sim 10^{-5}$, (ii) for white dwarfs : $\frac{M}{R}\sim 10^{-3}$, (iii) for neutron star :  $10^{-1}<\frac{M}{R}<\frac{1}{4}$, (iv) for ultra compact star : $\frac{1}{4}<\frac{M}{R}<\frac{1}{2}$ and (v) for Black hole :  $\frac{M}{R}=\frac{1}{2}$.
The surface redshift function $z_s$ for our model of compact star is obtained from the relationship,
\begin{eqnarray*}
1+z_s & = & \frac{1} {\sqrt{1-2u(R)}}, \nonumber\\
\Rightarrow~z_s&=&e^{\frac{ar^2}{2}}.
\end{eqnarray*}
To understand
strong physical interaction between particles inside the compact object
and its EoS, the surface redshift plays a dynamic role. In the absence of a cosmological
constant the surface redshift $z_s$ lies in the range
$z_s\leq 2$ \cite{buch,48,49}. On the other hand, in the presence
of a cosmological constant $\Lambda$, for an anisotropic star, the surface redshift obeys the
inequality $z_s \leq 5$ proposed by Bohmer and Harko \cite{49}. We have calculated the value of surface redshift for various compact stars in Table~2. It is also clear from the table that the value of surface redshift for these compact stars lies within the range $z_s<0.5$. \par
The gravitational redshift for the model of the compact star is obtained as,
\begin{eqnarray*}
  z &=& \frac{1}{A(r)}-1=\frac{a}{aD + C e^\frac{a r^2}{2}}-1,
\end{eqnarray*}
The central value of the gravitational redshift is obtained as,
\begin{eqnarray*}
  z_c &=& \frac{1}{A(r)}-1=\frac{a}{aD + C }-1,
\end{eqnarray*}
Now for a physically reasonable model, $z_c>0$, which gives, $a(1-D)>C.$
\begin{eqnarray}
\frac{dz}{dr}&=&-\frac{C a^2 e^{\frac{a r^2}{2}} r}{\Big(aD + (C e^{\frac{a r^2}{2}}\Big)^2}<0,
\end{eqnarray}
and at $r=0$, $\frac{d^2z}{dr^2}=-\frac{a^2 C}{(C + a D)^2}<0$.
It implies that gravitational redshift is monotonic decreasing function of $r$. The profile of the gravitational redshift is shown in fig~\ref{grav}.
\section{Generating function }\label{sec10}
Lake \cite{lake1} proposed an algorithm which generates all regular static spherically symmetric perfect-fluid solutions of Einstein's equations. Herrera {\em et al.} \cite{her08} extended this work by introducing locally anisotropic fluids and proved that two functions instead of one is required to generate all possible
solutions for anisotropic fluid. Very recently Ivanov \cite{iva20} also obtained the generating functions based on the condition for the existence
of conformal motion (conformal flatness in particular) and the Karmarkar's \cite{kar} condition for
embedding class one metrics.
Now by introducing DB \cite{db} transformation \[x=r^2,~~~ V(x)=\frac{1}{B^2},~~\text{ and}~~~ y(x)=A^2,\]
and using the notation, $\Delta=p_t-p_r$, from eqns.(\ref{8}) and (\ref{9}),we get,
\begin{eqnarray}
  \frac{dV}{dx}\Big(1+x\frac{\dot{y}}{y}\Big)+V\Big[\Big(2\frac{\ddot{y}}{y}-\frac{\dot{y}^2}{y^2}\Big)x-\frac{1}{x} \Big]=\kappa \Delta-\frac{1}{x},
\end{eqnarray}
The above equation can be denoted as,
\begin{eqnarray}
\frac{dV}{dx}+R(x)V=S(x),
\end{eqnarray}
which is linear equation of $x$.
The integrating factor of the above equation is, \[e^{\int R(x) dx}\]
and the solution of the above equation is,
\[V(x)=e^{-\int R(x) dx}\int S(x)e^{\int R(x) dx} dx +k\]
where $k$ is the constant of integration and
\begin{eqnarray}
R(x)&=&\frac{\Big(2\frac{\ddot{y}}{y}-\frac{\dot{y}^2}{y^2}\Big)x-\frac{1}{x}}{1+x\frac{\dot{y}}{y}},\label{p1}\\
S(x)&=&\frac{\kappa \Delta-\frac{1}{x}}{1+x\frac{\dot{y}}{y}}\label{p2}.
\end{eqnarray}
From the above discussion it is clear that the model of the compact star stands on two functions $R(x)$ and $S(x)$, where as $R(x)$ and $S(x)$ depends on $y(x)$ and $\Delta$. Therefore for our model the two generating functions are,
\begin{eqnarray}
y(x)&=&\frac{\Big(aD+C e^{\frac{a x}{2}}\Big)^2}{a^2},\\
\Delta &=& \frac{1 - e^{-ax} (1 + ax)}{x}.
\end{eqnarray}

\section{Discussion and concluding remarks}\label{sec11}

In our present paper, We have successfully obtained a new model of anisotropic compact star in (3+1)-dimensional spacetime  using Krori-Barua (KB) ansatz \cite{kb}. To solve the field equations, we have assumed $g_{rr}$ metric potential and a physically reasonable choice of the anisotropic factor $\Delta$ and the remaining physical parameters like $g_{tt},\,p_r,\,p_t$ have been determined by solving it. We have shown that the physical quantities $\rho,\,p_r,\,p_t,\,V_r^2,\,z$  are monotonically decreasing function of $r$ from the center of the star to the boundary. Proposed model does not suffer from any kind of singularities. One can see that the both the metric potentials, $\Gamma$, mass function, compactness as well as the surface redshift all are increasing function with increase in radius. The equilibrium condition of the present model has been discussed with the help of the TOV equation. From fig.~\ref{tov}, we see that among the three forces, gravitational force is attractive in nature and the other two forces are repulsive. Between anisotropic and hydrostatic forces, hydrostatics forces always dominates the anisotropic force and the effect of gravitational force is dominating among all and it is counterbalanced by the combine effect of $F_g$ and $F_a$ to help the system to maintain equilibrium.\par

We have calculated the values of $a,\,C,\,D$, central density, surface density, central pressure, compactness factor and surface redshift of some well known compact star candidates Vela X -1 with observational mass and radius $1.77 \pm 0.08$ and $9.56 \pm 0.08$ \cite{raw}, LMC X -4 with observational mass and radius $(1.04 \pm 0.09)M_{\odot}$ and ($8.301 \pm 0.2$)km. \cite{raw}, 4U 1608 - 52 with observational mass and radius $(1.74 \pm 0.14)M_{\odot}$ and $(9.528\pm 0.15)$ km. \cite{guv}, PSR J1614 - 2230 with observational mass and radius $(1.97 \pm 0.04)M_{\odot}$ and $(9.69 \pm 0.2)$ km. \cite{demo},  EXO 1785 - 248 with observational mass and radius $(1.3 \pm 0.2)M_{\odot}$ and $(8.849 \pm 0.4)$ km \cite{ozel} in table $1$ and $2$ and all the profiles are drawn for the compact star candidates PSR J1614 - 2230 4U 1608 - 52 considering the mass and radius ($1.97 M_{\odot}$ ,\,9.69 km) and ($1.65 M_{\odot}$ ,\,9.4 km) respectively. From the table~2, it can be seen that the calculated values of the central density, surface density and central pressure are respectively $\sim 10^{15}~gm.cm^{-3},\,\sim 10^{14}~gm.cm^{-3}$ and $\sim 10^{35}~dyne.cm^{-2}$ which are physically reasonable. Moreover twice the ratio of mass to the radius satisfies Buchdahl's bound $\frac{2M}{R}<\frac{8}{9}$ and all the energy conditions are satisfied as well. The stability of the present model is verified through causality condition, relativistic adiabatic index, Herrera's cracking method and Harrison-
Zeldovich-Novikov conditions and none of them is violated by the present model. Moreover to avoid the discontinuity of the tangential pressure at the boundary of the compact star we have matched our interior spacetime to the exterior vacuum solution in presence of thin shell.\par
From the above discussion, it can be concluded that our presented model of the compact star candidates fits very well with the observed values of masses and radii and therefore it may be used as a viable model to describe
strange stars in the background of general
relativity.\\

\bibliographystyle{unsrt}
\bibliography{kb20}

{\bf Acknowledgements} P.B is thankful to IUCAA, Government of India, for
providing visiting associateship.
\end{document}